\documentstyle[12pt]{article}
\input{psfig}
\begin{document}
\newcommand {\ber} {\begin{eqnarray*}}
\newcommand {\eer} {\end{eqnarray*}}
\newcommand {\bea} {\begin{eqnarray}}
\newcommand {\eea} {\end{eqnarray}}
\newcommand {\beq} {\begin{equation}}
\newcommand {\eeq} {\end{equation}}
\newcommand {\state} [1] {\mid \! \! {#1} \rangleg}
\newcommand {\sym} {$SY\! M_2\ $}
\newcommand {\eqref} [1] {(\ref {#1})}
\newcommand{\preprint}[1]{\begin{table}[t] 
           \begin{flushright}               
           \begin{large}{#1}\end{large}     
           \end{flushright}                 
           \end{table}}                     
\def\Acknowledgements{\bigskip  \bigskip {\begin{center} \begin{large}
             \bf ACKNOWLEDGMENTS \end{large}\end{center}}}

\newcommand{\half} {{1\over {\sqrt2}}}
\newcommand{\dx} {\partial _1}

\def\Dslash{\not{\hbox{\kern-4pt $D$}}}
\def\cmp#1{{\it Comm. Math. Phys.} {\bf #1}}\begin{titlepage}
\titlepage
\vskip 1cm
\centerline{{\Large \bf Small object limit of Casimir effect}}
\centerline{{\Large \bf and the sign of the Casimir force }}
\vskip 1cm

\centerline{O. Kenneth and  S. Nussinov}

\vskip 1cm
\begin{center}
\em School of Physics and Astronomy
\\Beverly and Raymond Sackler Faculty of Exact Sciences
\\Tel Aviv University, Ramat Aviv, 69978, Israel
\end{center}
\vskip 1cm
\begin{abstract}
We show a simple way of deriving the Casimir Polder interaction, present some general arguments on the finiteness and sign of mutual Casimir interactions and finally we derive a simple expression for Casimir radiation from small accelerated objects.  
\end{abstract}
\end{titlepage}

{\bf Introduction}\newline
The purpose of this short paper is threefold:
(a)First we show by using direct path integral techniques that the Casimir attraction\cite{1} between objects small relative to their separation does become the celebrated Casimir-Polder interaction\cite{2}.
(b) We comment on the sign of casimir forces when the type of the boundary condition is changed and for different geometries.
and finally in(c) we find a simple and very suggestive expression for the "Casimir radiation" of such small bodies when these are accelerated.
While many of the issues /techniques of this paper have been considered/used before\cite{3,4} we have some novel, hopefully useful insights and results.

\vspace{2mm}{\bf (a)Application to Van-der-Waals and Casimir-Polder}\newline
The macroscopic Casimir force between say dielectric (or conducting) plates can be obtained by summing up the pairwise microscopic casimir interactions between polarizable atoms\cite{6mil}. Because these pairs are not situated in vacuum multiple scattering corrections occur and are important. Here We would like to pursue the reverse path of deducing forces between microscopic objects by adopting the same path integral techniques that have been used for evaluation the Casimir force between macroscopic objects say the parallel plates force and also for the special variant with different directions of conduction\cite{a}.
 Consider the general case when
some field $\Phi$ is forced to vanish on some domain $\cal A$ . This condition may be enforced by adding to the action
a lagrange multiplier $\int_{\cal A}\Phi Jd^4x$. After integrating over ${\cal D}\Phi$ we are left with \beq\label{vdw}
{\cal Z}=\int {\cal D}J  \exp(-{i\over 2}\int\int_{{\cal A}\times{\cal
    A}}J(x)\Delta (x-x')J(x')d^4xd^4x') \eeq Where $\Delta$ is the
propagator of the field $\Phi$. In general $\Phi$ may be a vector and
$\Delta$ a matrix. 
Equation(\ref{vdw}) applies as well for
fields $\Phi$ which are not the basic canonical degrees of freedom provided that we use the correct propagator.
\footnote{Indeed the "field" $\Phi$ can be a composite field or some effective fluctuation in a real medium. So long as the distance scales in the problem (size of $\cal A$) are much larger then the compositeness scale (or lattice size) the field and its possible interactions can be described by an appropriate effective low energy Lagrangian. Only the lowest order quadratic terms yielding the infrared most relevant parts need to be maintained}

 For a static configuration we fourier transform in time
writing  \beq\label{sdw} {\cal Z}=\int {\cal D}J  \exp(-{i\over
  2}\int\int_{{\cal A}\times{\cal A}}J(x,\omega)\Delta
(x-x',\omega)J(x',-\omega)d^3xd^3x'{d\omega \over 2\pi}) \eeq Next
suppose that $\cal A $ is made of distinct bodies
 ${\cal A}=\bigcup {\cal A}_i$ which are small compared with the distances
between them. Then in the first approximation we may assume that each
body feels only the average value of $J(x)$ on each of the others. Denoting
 the distance between the i'th and j'th body by $r_{ij}$ we may write the
effective interaction between them as
 $\int J_i(\omega)\Delta_{\omega}(r_{ij})J_j(-\omega){d\omega\over 2\pi}$
where we define $J_i(\omega)=\int_{{\cal A}_i} J(x,\omega)d^3x
$. Computing the self interactions of each body is much more complicated and can be
done explicitly only for simple geometries. However we may, in the
spirit of renormalization assume that the self interaction is given by
some function of $J_i$ which, at least in the small field limit, may be approximated by some
quadratic expression which for obvious reasons we write as ${1\over 2\alpha_i}J_i^2$. By adding coupling to
an external field  $\delta S=\int_{\cal A}d^4x J\Phi_{ext}\simeq J\Phi$
and looking at the resulting equation of motion $J=\alpha\Phi$ we may
interpret $\alpha$ as some kind of susceptibility. In particular if
$\Phi$ is electric or magnetic field then $\alpha$ is just the usual electric or
magnetic polarizability. In this case J can be identified with the
electric or magnetic polarization. These arguments hold for
dielectrics as well as for conductors since the only difference would
be adding a "bare" term to the self interaction which we didn't compute
anyway. 
\newline Define a matrix $A_{ij}(\omega)$ by \beq\label{awd} 
 i\neq j\; 
A_{ij}(\omega)=\Delta_{\omega}(r_{ij}),
A_{ii}(\omega)={1\over\alpha} \eeq (If $\Phi$ is a vector then A is a
matrix in a larger space which we don't write explicitly) Then we obtain   \beq  {\cal Z}=\int
{\cal D}J(\omega) \exp (-i\int {d\omega\over 2\pi}\sum
J_i(\omega)A_{ij}(\omega)J_j(-\omega))\eeq    \beq E=i(\ln
{\cal Z})/T=-{i\over 2}\ln\det(...)=-{i\over 2}\int{d\omega\over 2\pi}\ln\det
A(\omega) \eeq We have assumed that the distances
$r_{ij}$ are large compared to the typical size $a$ of each
body. Under this assumption we necessarily have also
$A_{ii}>>A_{ij}$. ( if ${\Delta}(\omega,r)\sim{1\over r^n}$ then
$A_{ii}/A_{ij}\sim ({r\over a})^n$). Hence there is no reason not to use
this approximation also in the evaluation of $\det A(\omega)$. The leading term given by the product of the diagonal elements is a constant independent of the distances $r_{ij}$ which we may subtract. The next term, involving the diagonal $2\times 2$ minors yields:
  \beq \label{lihu} E=\sum_{i<j}{i\over
  4\pi}\int{d\omega}{A_{ij}(\omega)A_{ji}(\omega)\over
  A_{ii}(\omega)A_{jj}(\omega)} =\sum\int{i\alpha_i\alpha_jd\omega\over
  4\pi}\Delta_{ij}(\omega,r)^2\eeq 
In principle we could write extra correction to this expansion of the determinant however to be consistent we may not do that until we incorporated in our formalism the effect of higher multipole moments (and possibly of the nonlinear corrections if such exist).  
If $\Phi$ is the electric field $\vec E$
then \beq\label{efp}\hspace{-6mm} \Delta_{ij}(\omega,r)=\langle
E_i(0)E_j(r)\rangle=(\delta_{ij}({{\omega}^2\over r}-{i|\omega |\over
  r^2}-{1\over r^3})+{x_ix_j\over r^3}(-{\omega}^2+{3i|\omega |\over
  r}+{3\over r^2}))e^{-i|\omega|(r-i\varepsilon )}-{4\pi\over
  3}\delta_{ij}\delta^{(3)}(r) \eeq (this can be obtained by
differentiating  $\langle A_{\mu}A_{\nu}\rangle =- {g_{\mu\nu}\over
  r}e^{-i|\omega|(r-i\varepsilon)}$) and the usual Casimir-Polder
energy is given by $i\alpha_1\alpha_2\int{d\omega\over
  4\pi}{\Delta_{ij}(\omega,r)}^2=-{23\over 4\pi}{{\alpha}_1{\alpha}_2\over r^7}$. (The short distance Van der Waals relation can also be obtained from (\ref{lihu}) by letting $\alpha=\alpha(\omega)$ and assuming that the maximal frequency for which $\alpha$ is non-negligible is small compared with $1\over r$)
The calculation still holds when different fields vanish on different
bodies. For instance the interaction energy between electric and
magnetic objects with polarizabilities ${\alpha}_E,{\alpha}_B$ will be by
exactly the same reasoning \beq {i{\alpha}_E{\alpha}_B\over 4\pi}\int
d\omega{\langle B_i(0)E_j(r,\omega )\rangle}^2={i{\alpha}_E{\alpha}_B\over 4\pi}\int
d\omega{\left(i\omega({-i|\omega|\over r}-{1\over r^2}){{\varepsilon}_{ikj}x_k\over r}e^{-i|\omega|r}\right )}^2={7{\alpha}_E{\alpha}_B\over 4\pi}{1\over r^7}\eeq And (assuming $\alpha>0$) this force is repulsive.

The general result(\ref{lihu}) valid for $r\gg a$ may be considered as the leading term (corresponding to the diagram\put(14,3){\circle{14}}\put(7,3){\circle*{3}}\put(21,3){\circle*{3}}\hspace{7mm} )in an $a\over r$ expansion. Corrections to it may originate from(1)higher multipole (i.e. higher derivative) interaction (2)possible nonlinear effects (3)further expansion of $\ln\det A(\omega)$. 
The last of these factors has some special interest in that it is the one responsible to multibody forces. Thus the leading three-body interaction corresponding to the diagram \put(0,-4){\circle*{3}}\put(0,-4){\line(2,3){10}}\put(10,11){\circle*{3}}\put(0,-4){\line(1,0){20}}\put(20,-4){\circle*{3}}\put(20,-4){\line(-2,3){10}}\hspace{5.5mm} is just the next term:
$$\delta E_{3b}=-\sum\int{id\omega\over 4\pi}{\it Tr}\left ( \alpha_i\Delta_{ij}\alpha_j\Delta_{jk}\alpha_k\Delta_{ki}+ \alpha_i\Delta_{ik}\alpha_k\Delta_{kj}\alpha_j\Delta_{ji}\right)$$
Here the trace is over internal indices of the field and the summation is over unordered triplets \{i,j,k\}.
For a scalar field we thus easily obtain $\delta E_{3b}={\alpha_1\alpha_2\alpha_3 \over \pi r_{12}r_{23}r_{31}(r_{12}+r_{23}+r_{31})}$. For the electric case we find using (\ref{efp}) a similar expression, the explicit formula for which is however too long to give any extra insight.
It should be remarked that the (2-body) corrections to eq(\ref{lihu}) due to higher derivative terms are typically of relative order $({a\over r})^2$ (since they originate from self interactions $\sim(\partial J)^2$) This is bigger then the  $({a\over r})^3$  3-body correction to the electric Casimir Polder force and smaller then the $a\over r$ 3-body correction to the scalar Casimir Polder force.

\vspace{2mm}{\bf (b)Finiteness and sign of the Casimir effect}\newline
The repulsion between systems with non vanishing electrical and magnetical polarizabilities come about in the above discussion rather simply it is just reflects the $i$ factor that the euclidian correlation function $\langle EB\rangle_{Euc}$ and similarly $\langle\phi\dot{\phi}\rangle_{Euc}$ pick relative to $\langle EE\rangle_{Euc},\langle\phi\phi\rangle_{Euc}$ (or  $\langle BB\rangle_{Euc}$) upon Wick rotation.\footnote{The relative repulsion of electrically and magnetically conducting plates and the intuitive explanation of this pattern have been discussed at length by\cite{s}}
The issue of the sign of the Casimir forces and some possible puzzles and paradoxes there have the subject of considerable interest. Here we would like to clear up this matter as much as possible.
While we will make in the following many heuristic remarks we will clearly separate those from rigorous parts

 The electric Casimir Polder force between two conductors small compared to their mutual distance is attractive as is the force between two atoms. Therefore it is natural to expect that the total Casimir force between any two conducting objects  should also be attractive. However because of the non additive nature of the Casimir force (and specifically the existence of many body and not only 2-body interactions)this conclusion is not quite obvious.
Indeed for the case of two conducting plates adding the Casimir Polder interactions (and taking the formal limit $\epsilon\rightarrow\infty$ in the Clausius Mossotti relation ${4\pi n\alpha\over 3}\simeq{\epsilon-1\over\epsilon+1}$ with $n$ the molecular density and $\alpha$ the polarizability) falls short of the correct answer ${W\over A}={\pi^2\over 720}{\hbar c\over a^3}$ by$\simeq 30\%$\cite{6mil}. The gap is bridged by many body interactions\footnote{The approximations involved in deriving the Clausius Mossotti also have some affect. However these approximations are also related to the many body interactions}. Could it be that for some special geometries we can even achieve a complete sign reversal and repulsion between, say, two electrically conducting bodies?

The issue of the sign of the casimir energy cannot be meaningfully addressed unless we find for it a convergent, well defined expressions. Let us then consider the partition function (\ref{sdw}) for the case of two conductors $\Sigma_1,\Sigma_2$ whose centroids are at a distance $a$. Further assume that Wick rotations have been made so that we can use the euclidian form of $\cal Z$ and of the propagator ($\sim{1\over(x-y)^2}$ for 4d).
Let us next divide out ${\cal Z}(a)$ by  ${\cal Z}(a\rightarrow\infty)$ which is equivalent to subtracting the Casimir self energies\footnote{Each of the Casimir energies separately is quartically divergent} of $\Sigma_1,\Sigma_2$:
\beq\label{uu}{e^{-E(a)T}\over e^{-E(\infty)T}}={\int\prod dJ^{(1)}(x)\prod dJ^{(2)}(y)e^{-\int{J^{(1)}(x)J^{(1)}(x')\over(x-x')^2}}e^{-\int{J^{(1)}(x)J^{(2)}(y)\over(x-y)^2}}
e^{-\int{J^{(2)}(y)J^{(2)}(y')\over(y-y')^2}}\over{\int\prod dJ^{(1)}(x)\prod dJ^{(2)}(y)e^{-\int{J^{(1)}(x)J^{(1)}(x')\over(x-x')^2}}e^{-\int{J^{(2)}(y)J^{(2)}(y')\over(y-y')^2}}}}\eeq

In this expression the infinite contributions ${J^{(1)}(x)J(x')^{(1)}\over(x-x')^2}\;\;,x'\rightarrow x$ for $x,x'\in\Sigma_1$ etc divide out and eq(\ref{uu}) is finite and well defined. In particular the coefficient $1\over(x-x')^2$ of the mixed product$J^{(1)}(x)J(x')^{(2)}$ is bound by $1\over a^2$ with $a=\min|x-y|,x\in\Sigma_1,y\in\Sigma_2$, the minimal distance between the conductors and is finite.
In particular we write $\Delta ={1\over 4\pi^2}{1\over (x_1-x_2)^2+(y_1-y_2)^2+(z_1-z_2)^2+ (t_1-t_2)^2+\delta^2}$ with an infinitesimal $\delta$ introduced.
This $\delta$ will not affect the ratio in eq(\ref{uu})

The basic difference between the problems of two disjoint objects and the single object problem is that the first is indeed finite and needs no renormalization\footnote{In the context of finite temperature this point was recently elaborated by\cite{rmf}}(though some regulator may help in its evaluation).
This fact is physically obvious since changing the relative placement of two distinct object does not change the relative position of points belonging to the same object. Thus only pairs of points whose mutual distance is greater then $a$ change relative position and there can be no UV divergence.
On the other hand when considering changing the radius of a single sphere, say, we are considering a process in which any pair of points including infinitesimally closed pairs change relative position  and thus it is hardly surprising that the UV divergence are much more severe.
In the path integral approach, which we use here, these difficulties manifest themselves in ambiguity as to how to relate the measures ${\cal D}J$ for different radii since these are in fact measures on different spaces.

Physically the problem is to identify the bulk energy term of the sphere say, which is supposed to be subtracted in order to leave only the casimir part. If we think of the sphere as representing a real conductor then this bulk term is supposed to represent (at least part of) the binding (and surface) energy of the atoms making up the conductor. It is well known that this binding energy is a result of residual electric interactions just like the van der Waals and the Casimir interactions. Therefore it is not surprising that we have a difficulty in finding a natural way of separating them.

It is often stated\cite{5,em} that the Casimir force between two close by conducting hemispheres is repulsive. This assertion is based on the positivity of the known result for the Casimir energy of a spherical conducting shell.
Much effort and calculational skills have been devoted into the evaluation of the casimir energies for conducting bodies of various shapes. There it was found that for cylinder and parallelpipes of sufficiently large aspect ratio the casimir energy is negative and becomes positive for a closed cylinder/parallelpipe  of smaller aspect ratio as the more spherical geometry is approached.
Clearly no simple rule pertaining to curvature and/or topology can be used for the sign and one needs to compute each case separately\cite{8}.\footnote{In the discussion of eigenmodes inside cavities there is a well defined systematic hierarchy of volume surface,curvature etc terms which corresponds to the quartic etc divergences}

It is conceivable that in Kaluza Klein models with compactified dimensions the Casimir energy density can provide positive or negative cosmological constant.

However in the electromagnetic 3+1dim context we believe that the values and signs of Casimir energies for the various closed shapes are at best of academic interest only
and irrelevant for the force between different bodies. In all cases where the sign of the effect can be physically addressed by changing the relative distance between two objects, which are mainly electrically or mainly magneticly polarizable, the Casimir force will be attractive. 

In particular we note that we cannot infer repulsion between two hemispheres from the positive Casimir energy for the sphere.
The problem of a single conducting shell and the problem of two close by conducting hemispheres are completely different. Based on the most naive physical intuition one should expect that the force between the two hemispheres depends on what is happening in the area where they are the closest to each other i.e. near the edges. But clearly the edges are just the place where the two hemispheres configuration is most drastically different from that of the spherical shell.
Indeed let us consider two infinitesimal yet ideally conducting equatorial "rims" at the boundaries of the hemispheres. The wavelength of modes relevant to the interaction between these rims is $\lambda\sim\delta$ with $\delta$ the separation between the rims. Hence we may, in first approximation, neglect most of the hemispheres consisting of the more distant parts in evaluating the rim-rim interaction.
The interaction between two rims in vacuum is readily evaluated (up to $\ln(\delta/d)$ correction with $d$ the rim thickness) to be $F=\#_12\pi R\hbar c/\delta^3$. This attraction exceeds $F_{sphere}=\#_2{\hbar c\over R^2}$  with $\#_{1,2}$ some numerical coefficients by vast $(R/\delta)^3$ factor. Consequently the net force between two hemispheres remain attractive.

Some heuristic argument in favour of the tendendcy for repulsion amongst the two hemispheres, follows from considering the surface charge on a conductor placed in an electric field and the affect of the extra field generated by this induced charge. It is easy to see that the normal component of this extra field will be directed outward or inward depending on whether the conductor is concave or convex. As a result one may expect collective phenomena to increase the attractive force between convex bodies and to decrease the force between concave bodies (though realistic bodies cannot of course be globally concave).The fact that we ignored retardation in this argument is not a serious drawback since the argument works in four-dimensions as well as in three.

In spite of this reasoning we find it hard to believe that the force can actually change sign. We therefore conjecture (counter to common belief) that the electric Casimir force between two conducting objects of any shape is always attractive. Similarly we conjecture the same assertion holds for any two objects on which a real scalar field vanishes. 
    
We next proceed to a more formal discussion of our conjectured universally attractive casimir forces between objects of similar electric properties (say both with dielectric coefficients $\epsilon$ or both conducting\footnote{the form of the casimir Polder interaction suggests that also bodies with both electric and magnetic polarizabilities attract so long as the average $\alpha_E^{(i)}/\alpha_B^{(i)}$ ratios are similar})

We can show that the energy of two conductors placed at any finite distance from each other is always lower then when their distance tends to infinity. It is easiest to keep track of the signs by expressing the energy using the euclidian functional integral as
$e^{-ET}={\cal Z}=\int{\cal D}J\exp(-J*\Delta *J)$ With $J$ current living only on the volume of the two objects. If we denote by $J_1,J_2$ the current in the first and second conductor then by making a change of variables $(J_1,J_2)\rightarrow(J_1,-J_2)$  in the functional integral we can change the sign of the term $J_1*\Delta *J_2$ and by adding the two equivalent expressions for $\cal Z$ we find:
\beq  e^{-ET}={\cal Z}=\int{\cal D}J_1{\cal D}J_2\exp(-J_1*\Delta *J_1-J_2*\Delta *J_2)\cosh(J_1*\Delta *J_2)\eeq 
Now the dependence of this on the relative position of the two objects is only through the terms $\cosh(J_1*\Delta *J_2)=\cosh\int d^3xd^3y{d\omega\over 2\pi}J_1(x,-\omega)\Delta (x-y,\omega)J_2(y,\omega)$ note that although $J(\omega)$ unlike $J(t)$ is not real the combination $J_1(-\omega)J_2(\omega)+J_1(\omega)J_2(-\omega)$ which multiply $\Delta (x-y,\omega)$ is real (reality would be obvious if we use sine and cosine instead of exponential fourier transform which may be a better choice when one wants to keep track of the signs).
 Since $\cosh$ of a real argument is always$\geq 1$ we immediately conclude that Casimir energy of two objects at a finite distance from each other is always smaller then when their distance tends to infinity. Hence as long as the two objects are far they necessarily attract.

To prove however that the two objects keep attracting at all distances we need to show that the derivative ${\partial\over \partial a}E(a)$ of the energy as a function of the relative separation does not change sign.
We will next attempt to motivate this much less obvious feature. To this end let us revert to the original Euclidian pure configural space expression for the interaction energy with $\Delta ={1\over 4\pi^2}{1\over (x_1-x_2)^2+(y_1-y_2)^2+(z_1-z_2)^2+ (t_1-t_2)^2+\delta^2}$ where an infinitesimal $\delta$ was introduced to avoid infinite self interactions. Approximating the volumes (or surfaces) of the two objects by finite numbers $N_1,N_2$ of points we have:
\hspace*{-3mm}\bea \lefteqn{e^{-E(a)T}=\int\prod dJ_l\exp\left\{ -{1\over 2}\sum_{l=1}^{N_1}\sum_{m=1}^{N_1}J_l^{(1)}{1\over(x_l-x_m)^2+(z_l-z_m)^2+\delta^2}J_m^{(1)}-\right.} \\ & \left.-{1\over 2} \sum_{l=1}^{N_2}\sum_{m=1}^{N_2}J_l^{(2)}{1\over(x'_l-x'_m)^2+(z'_l-z'_m)^2+\delta^2}J_m^{(2)}-\sum_{l=1}^{N_1}\sum_{m=1}^{N_2}J_l^{(1)}{1\over(x_l-x'_m)^2+(z_l-z'_m+a)^2}J_m^{(2)}\right\}\nonumber\eea
and likewise for $e^{-E(\infty)T}$. With $J_l^{(1)},J_m^{(2)}$ the currents at the $N_1,N_2$ points $(\vec{x}_l,z_l)$ and  $(\vec{x}'_l,z'_l)$ on each of the conductors respectively. We envision separation along the $z$-axis and $z_m,z'_m$ are the distances along the $\pm z$ axis of the points from two initial planes parallel to the $x,y$ planes (so that both $z_l\geq 0,z'_l\geq 0$). We use $\vec{x}_l=(x_l,y_l,t_l)$ to indicate the other coordinates.
$a$ is the relative displacement along the $z$-axis of the two planes (or objects) from the "initial" tangency point.

Taking the $a$ derivative we find
\beq\label{dtr}{\partial\over\partial a}e^{-ET}=\int\prod dJ_l \sum_{l=1}^{N_1}\sum_{m=1}^{N_2}J_l^{(1)}{2(z'_l+z_m+a)\over((x_l-x'_m)^2+(z_l-z'_m+a)^2)^2}J_m^{(2)}\exp{\left\{........\right\} }\eeq
The above expression breaks into $N_1N_2$ contributions corresponding to specific choices of $l=l^0,m=m^0$. Thanks to the $\exp(-J^{(1)}{1\over(.....)}J^{(2)})$ factor we have in each term separately the regions in the $J_{l^0},J_{m^0}$ plane  where the signs of $J_{l^0}$ and $J_{m^0}$ are opposite yield a larger contribution to the $dJ_{l^0}^{(1)}dJ_{m^0}^{(2)}$ integral (in absolute magnitude). Because of the positivity of $2(z'_l+z_m+a)$ these "dominant" regions make then consistently a negative contribution to ${\partial\over\partial a}e^{-ET}$.
Since we have only  $N_1+N_2$  independent $dJ^{(1)}dJ^{(2)}$ integrations rather then $N_1N_2$ independent integrations on all $J_{l^0}^{(1)}J_{m^0}^{(2)}$ pairs we cannot utilize the above to rigorously deduce that the expression in (\ref{dtr}) above is negative. However since there is no obstruction to choosing also several $J_l^{(1)}$ to be of opposite sign from several of $J_m^{(2)}$, this is very strongly suggestive of our claimed monotonic behaviour of $E(a)$ and the negative $F_{casimir}$.

The above arguments can be readily extended to the vector case and also to dielectrics.

 The $\sum J_\mu^{(i)}A^\mu_{(i)}$ lagrange multiplier form suggests that we identify the $J$'s with induced conserved electric currents.
The opposite sign of these conforms to the heuristic argument that in the two conducting plates say, the attraction is due to charge patches and image charge of the opposite sign on the other plate. It also allows us to understand why there is still residual attraction between two parallel plates which conduct in perpendicular directions. While the spatial parts of $J_1$ and $J_2$ are then orthogonal and do not contribute, the temporal $J_1^0$ and  $J_2^0$ are not.

Coming back to the starting point of the present section
 we use the above reasoning  to find the affect of other boundary conditions.
 The simplest example in which this can be seen is when we demand that on one object the field $\phi$ vanishes while on the other the time derivative $\dot{\phi}$ is zero. In this case most of the above reasoning goes the same way. The only important difference is that the term representing the interaction between the two objects is built out of the two point function $\langle\phi\dot{\phi}\rangle\sim i\omega {1\over r}e^{-r\sqrt{\omega^2+m^2}}$ which is pure imaginary instead of the real function $\langle\phi{\phi}\rangle\sim{1\over r}e^{-r\sqrt{\omega^2+m^2}}$. This has the effect of transforming the hyperbolic cosine into a trigonometric cosine. Now since $\cos$ unlike $\cosh$ is not a monotonically increasing function of (the absolute value of) its argument we cannot complete the argument to deduce that the force (at least at large distances) is attractive. In fact since for small argument the function $cos$ is decreasing we may conclude from the above that the force will be repulsive provided that the mutual interaction is small. This conclusion should be valid whenever the distance between the the objects is large enough.
By exactly the same argument it follows that the force between an electric and a magnetic object is repulsive (at least for large distances) simply because the euclidian correlation function $\langle EB\rangle(r,\omega)$ is pure imaginary.

We remark that the imaginarity of $\langle\phi\dot{\phi}\rangle$,$\langle EB\rangle$ is only a result of the i factor which
 $\dot{\phi}$ and $E$ picked up upon the wick rotation. Had the euclidian action including the lagrange multipliers terms been real we would have necessarily obtained real $J_1$-$J_2$ interaction and hence an attractive force (at least at large distances).

There is one more well known example of a repulsive Casimir force we have not addressed so far. This is the case of a scalar satisfying Dirichlet boundary conditions on one plate and Neumann condition on another. (For a massless scalar in 1+1 dimensions the repulsive nature of the force follows immediately from the $\zeta$-function regularized sum $\sum(n+{1\over 2})={1\over 24}$ which is positive contrary to $\sum n=-{1\over 12}<0$ for Dirichlet- Dirichlet boundary conditions a rather unintuitive argument).
Our previous computations might suggest trying to enforce the Neumann condition by adding a term of the form $J\partial_z\phi$ to the action. This however is wrong since the Neumann boundary condition is not equivalent to demanding $\partial_z\phi=0$ over the plate. When the Neumann boundary condition is said to be satisfied on a plate it is meant that the field on one side of it is completely independent of the field on the other side while demanding $\partial_z\phi=0$ enforces the field to have exactly the same value on both sides. Indeed demanding $\partial_z\phi=0$ (and $\phi=0$ on the other plate) would yield an attractive force (dependent on the first plate width d and explicitly given by\footnote{ In evaluating the self interaction we assumed $J(z)$ to vary over the width of the plate in such a way that $\int \partial_z^2D(z_1-z_2)J(z_2)dz_2$ is constant over it, this forced $J(z)$ to be a (specific) second degree polynomial in $z$. This computation is in fact correct only for $d<<a$ and therefore the nonlinear terms in $d$ may be omitted.}
$E={1\over 2}\int{d^3k\over(2\pi)^3}\log(1-{1\over 2}kde^{-2ka})$ ). The correct way to enforce the ordinary Neumann condition is by making the kinetic term $(\partial\phi)^2$ weight in the action tend to zero (rather then infinity) inside the body i.e. we may write the action in the form $S=\int\varepsilon(x)(\partial\phi)^2d^4x$ where $\varepsilon(x)$ vanishes inside the Neumann objects and is equal to 1 outside it. Enforcing Dirichlet condition may be done similarly by adding a term $\int_{\cal A}\infty\times\phi^2$ however since the specific mode $k=0$ is not too important we may as well use an extra term of the form $\int_{\cal A}\infty\times(\partial\phi)^2$ i.e. we may take $\varepsilon(Dirichlet)=\infty,\varepsilon(Neumann)=0$. Using this point of view the Dirichlet -Neumann repulsion is similar to the repulsion between para and dia-magnetic plates which is intuitively obvious.

\vspace{2mm}{\bf (c)Casimir radiation}\newline We next proceed to the last topic of this paper namely casimir type radiation.\footnote{Such issues have been discussed in the past by several authours see e.g.\cite{br}}
Time dependent boundary conditions will in general result in radiation even from neutral objects. The leading contribution is from pair production process. The matrix element  $\langle 0|k_1,k_2\rangle$ for it is essentially  $\langle JJ\rangle$-the correlator of the current we introduced as lagrange multipliers. Indeed considering for simplicity a scalar field and repeating the standard steps leading to (\ref{vdw}) one easily finds  $\langle{\Phi}_i{\Phi}_j\rangle-\langle{\Phi}_i{\Phi}_j\rangle_0= \Delta_{ik}\Delta_{jl}\langle
J_kJ_l\rangle$. Where $\langle..\rangle_0$ denotes the expectation value in the absence of the extra boundary conditions over $\cal A$.  
 Here $J$ should be
interpreted as a function $J(x,t)$ which is defined all over spacetime
but vanishes identically outside the region $\cal A$. Hence we have  $\langle 0|k_1,k_2\rangle=\int_{{\cal A}\times {\cal A}}d^4x d^4ye^{ik_1\cdot x+ik_2\cdot y}\langle J(x)J(y)\rangle$.

 Radiation is
related to the asymptotic behaviour at large distances. At such
distances every object will look approximately pointlike Hence it is natural to consider in the following radiation from an accelerating small, point like object. We will assume that the typical size $a$ of
this object is small compared with the wavelength of the emitted
radiation and compared with its typical acceleration time. The first
assumption allows us to assume that our $J$ variable can be regarded
as constant over the volume of the emitting object so that it is a
function $J=J(\tau)$ of only the self time $\tau$ along the world line of
the emitting object.The second assumption imply that the self
interaction of $J$ is not effected (to first approximation) by the
fact that our object is not static. Thus in the leading approximation $S\simeq{1\over 2\alpha}\int J(\tau)^2d\tau$ with the same $\alpha$ as in eq(\ref{awd}-\ref{lihu}) and $\langle J(\tau_1)J(\tau_2)\rangle\simeq\alpha\delta(\tau_1-\tau_2)$. Denoting the emmiting object trajectory by $x(\tau)$ we see that if a scalar field is forced to vanish on it the amplitude for particle pair creation is simpliy given by:
 \beq\langle 0|k_1,k_2\rangle
=\alpha\int d\tau e^{i(k_1+k_2)x(\tau)}\eeq
For electromagnetic Casimir radiation most of the the computation proceeds along the same lines. The only difference (except for extra indices) is the appearance of the extra factor $\omega_1\omega_2$ as a result the need to relate the field $\vec{E}$ which actually vanishes on $\cal A$ to the field $A_\mu$.
Thus we obtain:
\beq\langle 0|k_1\vec\varepsilon_1,k_2\vec\varepsilon_2\rangle
=\omega_1\omega_2\vec\varepsilon_1\cdot\put(0,8){\vector(1,0){11}}\put(5,8){\vector(-1,0){7}}\alpha\cdot\vec\varepsilon_2\int d\tau e^{i(k_1+k_2)x(\tau)}\eeq 

For a particle moving in a circular motion of radius $r$ and frequency $\omega$ $x=r\sin(\omega\tau),y=r\cos(\omega\tau),t=\tau/\gamma={\tau\over\sqrt{1-r^2\omega^2}}$
the integral giving the amplitude will turn into\newline  $\int d\tau\exp(ik_{\|}r\sin(\omega\tau)-i\gamma k_t\tau)=2\pi\sum_nJ_n(k_{\|}r)\delta({\gamma k_t}-n\omega)$ Where $k=k_1+k_2$ is the sum of the two emitted particles four momentum and $k_\|$ is its component in the plane of rotation.

The energy radiated per unit of time will therefore be 
\beq W=\sum_n{\alpha^2n\omega\over\gamma^3}{1\over (2\pi)^3}\int{d^3k_1\over 2k_1}{d^3k_2\over 2k_2}J_n(k_{\|}r)^2\delta(k_t-{n\omega/\gamma})A \eeq
where for a massless scalar $A=1$ while for the electromagnetic case $A=k_1^2k_2^2+(\vec{k}_1\cdot\vec{k}_2)^2$.
Assuming $\omega r<<1$ we find for the scalar $W={\alpha^2r^2\omega^6\over 30\pi}$ and for the electromagnetic problem \beq\label{wr} W_{el}={\alpha^2r^2\omega^{10}\over 567\pi}\eeq
The actual amount of radiation emitted under any normal circumstances is very tiny. To estimate $W$ consider the limit $a=$radius of spherical shell $a\simeq r$ and when the sphere moves around the circle with the speed of light
\footnote{formally the derivation fails in this limit but we use it only to normalize the "boundary values"}. Eq(\ref{wr})implies that in this limit our rotating sphere acts as a "photon production machine" which by "churning" the vacuum generates one (actually$1\over 567$) photon pair of energy $\hbar\omega$ per rotation i.e. during time $T={2\pi\over\omega}$. For other -nonextreme -situations this should be reduced by $\beta^8({a\over r})^6$.
It is hard to envision acceleration of macroscopic bodies to speed higher then $\beta\approx\alpha=$speed of (valance)electrons, in which case emision rate is$<{\alpha^{14}\over 567}\approx 10^{-33}$ phtons per rotation.

\end{document}